\documentclass[preprint2]{aastex63}

\def\bfc{}

\def\kms{km~s$^{-1}$~}
\def\cm3{cm$^{-3}$~}

\usepackage{graphicx}
\usepackage{epsf}
\usepackage{natbib}
\usepackage{enumitem}
\usepackage{amsmath}

\newcommand{\simgt}{\ga}
\newcommand{\simlt}{\lower.5ex\hbox{$\; \buildrel < \over \sim \;$}}

\providecommand{\sorthelp}[1]{}

\begin{document}

\title{Supernova Shocks in Molecular Clouds: Velocity Distribution of Molecular Hydrogen}

\shorttitle{Supernova Shocks in Molecular Clouds}

\author[0000-0001-8362-4094]{William T. Reach}
\affil{Universities Space Research Association, MS 232-11, Moffett Field, CA 94035, USA}
\email{wreach@sofia.usra.edu}

\author{Le Ngoc Tram}
\affil{Universities Space Research Association, MS 232-11, Moffett Field, CA 94035, USA}

\author{Matthew Richter}
\affil{University of California, Davis, CA USA}

\author{Antoine Gusdorf}
\affil{Observatoire de Paris, \'Ecole normale sup\'erieure, Sorbonne Universit\'e, CNRS, LERMA, 75005 Paris, France}

\author{Curtis DeWitt}
\affil{Universities Space Research Association, MS 232-11, Moffett Field, CA 94035, USA}

\begin{abstract}

Supernovae from core-collapse of massive stars drive shocks into the molecular clouds from
which the stars formed.
Such shocks affect future
star formation from the molecular clouds,
and the fast-moving, dense gas with compressed magnetic fields is associated with
enhanced cosmic rays.
This paper presents new theoretical modeling, using the Paris-Durham shock model,
and new observations at high spectral resolution, 
using the Stratospheric Observatory for Infrared Astronomy (SOFIA),
of the H$_2$ S(5) pure rotational line from molecular shocks in 
the supernova remnant IC~443.
We generate MHD models for non-steady-state shocks driven by the pressure of the IC~443 blast wave into gas
of densities $10^3$ to $10^5$ cm$^{-3}$. We present the first detailed derivation of the shape of
the velocity profile for emission from H$_2$ lines behind such shocks, taking into account the
shock age, preshock density, and magnetic field. 
For preshock densities $10^3$--$10^5$ cm$^{-3}$,
the H$_2$ emission arises from layers that extend 0.01--0.0003 pc behind the shock, respectively.
The predicted shifts of line centers, and the line widths, of the H$_2$ lines range from 20--2, and 30--4 km~s$^{-1}$, 
respectively. 
The {\it a priori} models are compared to the observed line profiles, showing 
that clumps C and G can be explained by shocks into gas with density 10$^3$ to $2\times 10^4$ cm$^{-3}$ and strong magnetic fields. Two positions in clump B were observed. For clump B2 (a fainter region near clump B), the H$_2$ 
spectrum requires a J-type shock into moderate density ($\sim 10^2$ cm$^{-3}$) with the gas accelerated to 100 km~s$^{-1}$
from its pre-shock location. Clump B1 requires both a magnetic-dominated C-type shock (like for clumps C and G)
and a J-type shock (like for clump B2) to explain the highest observed velocities. 
The J-type shocks that produce high-velocity molecules may be 
locations where the magnetic field is nearly parallel to the shock velocity, which makes it impossible for a
C-type shock (with ions and neutrals separated) to form.

\clearpage

\end{abstract}


\section{Introduction}

When a supernova explodes close to (or within) a molecular cloud, shocks of high ram pressure and a wide range of speeds are driven through the gas, which spans a range of densities. The remnant of one such explosion is IC~443.
Our understanding of the conditions in the environment of the supernova that produced IC 443 remain somewhat speculative. The presence of denser-than-average-ISM gas in the vicinity of IC 443 has been long known, with relatively dense atomic gas toward the northeast (NE) 
\citep{giovanellihaynes79} and a molecular cloud toward the South \citep{dickman92}.
From optical spectra of filaments in the NE, a shock velocity 65--100 \kms and preshock density 10--20 \cm3 was inferred \citep{fesen80}.
From millimeter-wave spectra of a shocked clump in the southwest, 
a preshock density $\sim 3000$ \cm3 was inferred \citep{vandishoeck93}.
Even apart from the dense clumps,
best estimates of the preshock density differ by 2 orders of magnitude. While \citet{chevalier99} inferred that the forward shock progressed into a molecular cloud with inter clump density 15 \cm3, analysis of the X-ray spectral imaging from XMM-Newton in comparison to shock models led to a pre-shock density of 0.25 \cm3 \citep{troja08}. 

The stellar remnant of the progenitor of IC~443 is likely a fast-moving neutron star (CXOI J061705.3+222127) in the wind nebula (G189.22+2.90) found
 in {\it Chandra} X-ray images \citep{olbert01,gaensler06}.
That the neutron star is much closer to the
southern edge of the supernova remnant than the NE rim, which is so bright in visible light and and X-rays, is readily explained by 
faster expansion of the blast wave into lower-density material that was NE of the progenitor, combined with a
`kick' velocity that was imparted to the neutron star during the explosion, sending it southward. 
An upper limit to the
proper motion of the neutron star of 44 mas/yr \citep{swartz15} leads to a lower limit on the
time since explosion, if indeed the neutron star is the stellar remnant \citep[cf.][]{gaensler06}. 
The center of explosion is not known; the neutron star is presently about $14'$ from the center
of the overall SNR, but the explosion center was more likely near the center of the molecular ring
\citep{vandishoeck93} which is about $7'$ from the neutron star.
This leads to an age constraint $t>9,500$ yr. The presence of metal-rich ejecta, and the brightness of
the X-rays, support a young age. We will adopt $t\sim 10^4$ yr (the youngest age that agrees
with the proper motion limit) as the SNR age.

Shocks into dense clumps are of particular interest, because they affect the structure and chemistry of potential star-forming material, and they are a likely site of cosmic ray acceleration.
Such shocks lead to broadened molecular line emission and, in special circumstances, masers. 
To first order, the ram pressure into the clumps is expected to be the same as $p_{\rm ram}$ for the shocks into lower-density gas in the NE.
A clump in a typical molecular cloud may have density  $n_c\sim 3,000$ \cm3, and the present-day ram pressures of IC~443 would drive a shock into the clump at
$v_c\sim 10$ \kms. 
This shock speed appears inadequate to explain the widths of molecular lines observed in the clumps, which range from 20--100 \kms. We provide an explanation for this
discrepancy in \S\ref{sec:conditions}.

\begin{figure*}
\plotone{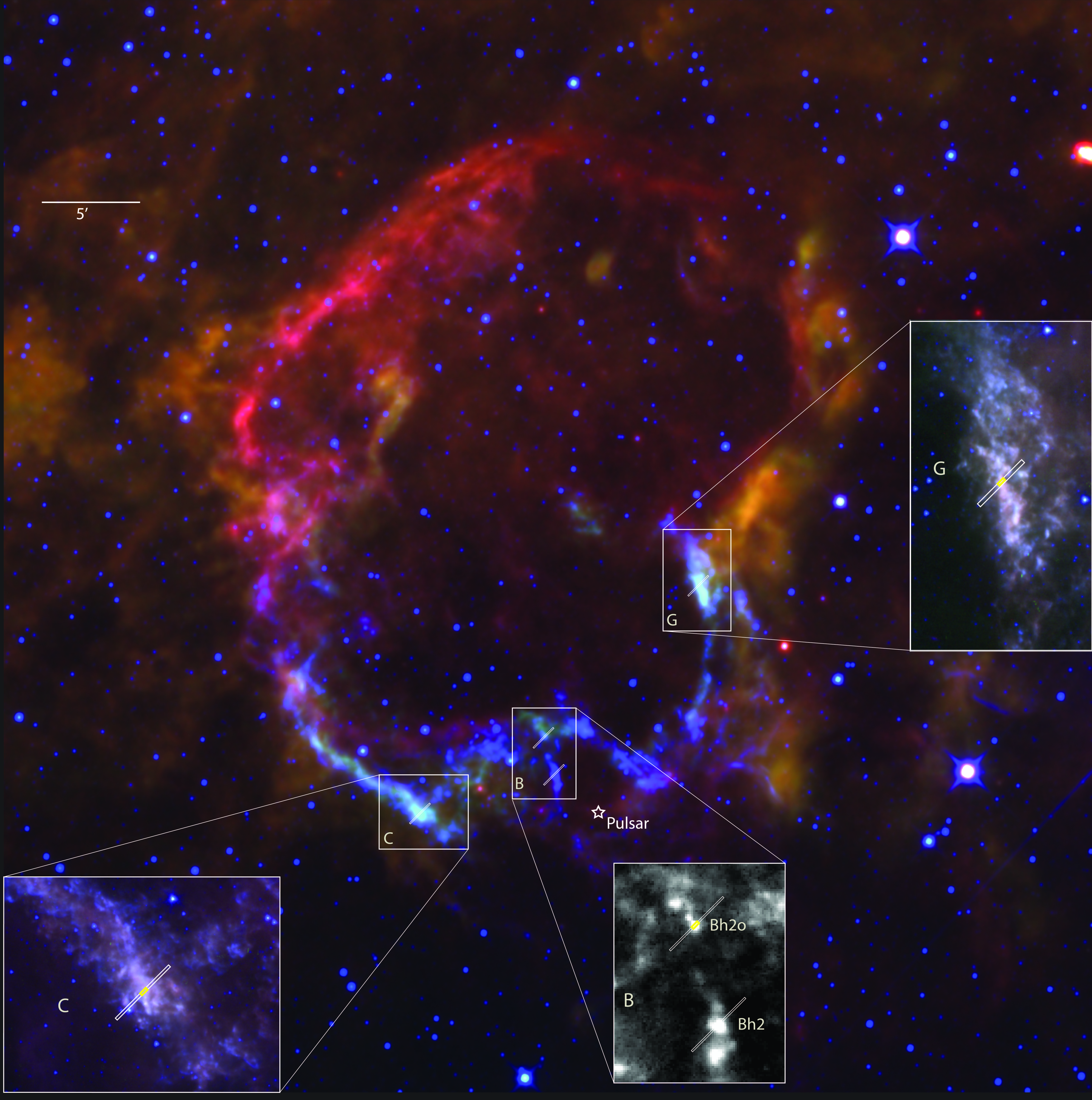}
\caption{{
{\it WISE} image of IC~443, with colors combined such that blue is band 1 (3.6 $\mu$m),
green is band 2 (4.5 $\mu$m), orange is band 3 (12 $\mu$m), and red is band 4 (22 $\mu$m) \citep[see][for the wavelength ranges of the broad bands]{wright10}.
Three regions (B, C, and G) in the $\omega$-shaped molecular region are shown with white boxes surrounding them.
Inset figures show archival {\it Spitzer} image enlargements of these three regions. For C and G, the insets are from a {\it Spitzer}/IRAC
\citep{fazioirac} with blue being band 1 (3.6 $\mu$m), green being band 2 (4.5 $\mu$m), and red being band 3 (5.8 $\mu$m). For region B (not covered by IRAC), the enlargement is the {\it Spitzer}/MIPS image at 24 $\mu$m in greyscale (white=bright).
For each target observed with SOFIA, the medium-resolution slit is shown as a white rectangle.
The present position of the neutron star is indicated with a $\star$ labeled `Pulsar'. 
The locations of the SOFIA/EXES medium-resolution slits are shown as white rectangles on the {\it WISE} image and {\it Spitzer} insets. The locations of the SOFIA/EXES high-resolution slits are shown as smaller, yellow rectangles on the {\it Spitzer} inserts.
The scale bar is $5'$ long, which corresponds to 2 pc at the assumed distance of IC~443.
}
\label{wise443}}
\end{figure*}

The importance of understanding shocks into dense clumps  increased significantly 
with detection of TeV emission \citep[MAGIC;][]{magic443}. 
The first resolved TeV \citep[VERITAS; ][]{veritas443} and high-energy $\gamma$-ray \citep[{\it Fermi}; ][]{fermi443} images of a supernova remnant were for IC~443, 
and they show a correspondence
in morphology with shocked molecular gas. Indeed, there is no correspondence between the TeV image and
the radio continuum, which would have been a natural expectation because the radio emission is also 
related to relativistic particles; however, the radio emission evidently traces shocks into lower-density gas, 
which do not lead to acceleration to TeV energies. 
The ability to pinpoint TeV emission to individual sources, and to individual sites within those sources, 
opens the exciting possibility of constraining the origin of cosmic rays by direct observation of environment
where they originate.
It is with this renewed emphasis on dense molecular shocks as the origin of 
gamma-ray emission from the interaction with particles accelerated to relativistic energies
in the earlier phases of the 
explosion that
we delve into the shock physics using new observations of H$_2$, by far the most abundant particle, from 
molecular shocks in IC~443.

\section{Molecular Shocks in IC 443\label{sec:conditions}}

The properties of the supernova remnant and the conditions of the shocks that are driven
into the interstellar gas are most readily understood in reference to an adiabatic explosion 
into a region of uniform preshock density. Variations in the in the density around the star are what
lead to the unique properties of each supernova remnant, but the uniform-density model is useful
for the properties of the lowest-density, highest-filling-factor medium. 
Consider a reference model of an explosion, with total energy $E$, in a uniform-density medium of preshock density $n_0$, occurring $t$ years ago, leading to a blast wave of radius $R$ expanding at rate $v$. 
\def\extra{In this case,
\begin{equation}
E = \frac{4\pi}{3}\mu m_{\rm H} n_0 R^3 v^2.
\end{equation}
}
For an adiabatic expansion (total energy $E_{51}$, in units of $10^{51}$ erg, remains constant), the solution is 
\begin{equation}
R = 12 
E_{51}^\frac{1}{5}
\left(\frac{n_0}{{\rm cm}^{-3}}\right)^{-\frac{1}{5}}
\left(\frac{t}{10^{4}\,{\rm yr}}\right)^\frac{2}{5}\,\,{\rm pc}
\end{equation}
and
\begin{equation}
v = 470 
E_{51}^\frac{1}{5}
\left(\frac{n_0}{{\rm cm}^{-3}}\right)^{-\frac{1}{5}}
\left(\frac{t}{10^{4}\,{\rm yr}}\right)^{-\frac{3}{5}}\,\,{\rm km~s}^{-1}.
\label{eq:vs}
\end{equation}
In IC~443, on the NE rim that is bright in radio and X-ray, 
the radius of the blast wave $R\simeq 8$ pc.
The X-ray analysis, taking into account the configuration of forward shock and reverse shock into the still-detectable, metal-enriched ejecta from the progenitor star, led to an age estimate of $t\simeq 4,000$ yr \citep{troja08}, while estimates based on optical observations led to an age estimate of $t\simeq 30,000$ yr \citep{chevalier99}. 
The ram pressure (per nucleon mass) into the present-day shocks
\begin{align} \label{eq:pram}
p_{\rm ram} \equiv n_0 v^2 & =
2.2\times 10^5
E_{51}^\frac{2}{5}
\left(\frac{n_0}{{\rm cm}^{-3}}\right)^\frac{3}{5} \\
& \left(\frac{t}{10^{4}\,{\rm yr}}\right)^{-\frac{6}{5}}\,\,
  {\rm cm}^{-3}\,({\rm km~s}^{-1})^2  \nonumber
\end{align}
in convenient units.

Using $n_0=0.25$ \cm3 and $t=4,000$ yr from the X-ray analysis \citep{troja08},
or using $n_0=15$ \cm3 and $t=30,000$ yr from \citet{chevalier99}, 
{\bfc and $E_{51}$=0.5} the same
$p_{\rm ram}\simeq 3\times 10^5$ \cm3 (\kms)$^2$ results. 
This equality arises because both models are matched to observable constraints that are related to 
the energy input into the interstellar material at the present time, which is directly tied to 
the ram pressure of the shock.
{\bfc As noted above, we adopt an intermediate age of 10,000 yr for IC~443, which is consistent with
the ram pressure if the preshock density is $n_{0}=2$ \cm3.}

Shocked molecular gas in IC~443 was first found from early CO(1-0) observations that showed 20 \kms linewidths toward some locations in the southern part of the SNR \citep{denoyer79}.
The ridge of shocked H$_2$ was imaged in the near-infrared 1--0 S(1) line by \citet{burton88}, who showed it was
``clear evidence for the presence of a shock, driven by the expanding gas of a supernova remnant, within a  molecular cloud." They also noted a lack of near-infrared Br$\gamma$, which limits the amount of ionized gas (and hence the prevalence of fast shocks). They  also explain the \ion{H}{1} 21-cm emission as the result of partial dissociation of H$_2$ in the
shocks where the blast wave interacts with a molecular cloud,
as opposed to preexisting stellar wind shells \citep[cf.][]{braun86}.
The important shock-diagnostic far-infrared line 63 $\mu$m line of \ion{O}{1} was detected
using the {\it Kuiper Airborne Observatory} and found to be well-correlated with the near-infrared H$_2$ \citep{burton90}.

The entire distribution of shocked H$_2$, and the clear distinction between the shocked gas in the NE and S parts
of the SNR were demonstrated with large-scale near-infrared imaging by 2MASS, where the NE portion of the SNR is much brighter in broad bands in the $J$ and $H$ bands, due to shocked \ion{Fe}{2}, and the S portion of the SNR is 
much brighter in the $K_s$ band, due to shocked H$_2$  \citep{rho01}. 
The {\it WISE} image of IC~443 shows this same distinction. In Fig.~\ref{wise443}, the NE portion of the SNR appears red, forming almost a hemispherical arc. The S portion appears blue, forming approximate the shape of the Greek
letter $\omega$. (The shape has also been described as the Roman letter W, or when incompletely imaged a letter S.)
The color variation can be understood as arising from line emission, similar to that seen by 2MASS but 
extended to the mid-infrared \citep{reach06}:
{\it WISE} channels 1 and 2 contain bright lines from highly-excited H$_2$ and CO, while channel 3 is dominated by
 PAH grains and \ion{Ne}{2} and \ion{Ne}{3} ions and channel 4 is dominated by small, solid grains and the 
 \ion{Fe}{2} ion. In this work we focus on the blue, $\omega$-shaped ridge.

The H$_2$ molecule is symmetric, so  dipole  transitions among the rotational levels (quantum number $J$) are forbidden by quantum selection rules. 
Quadrupole ($\Delta J=2$) transitions are permitted; they are spectroscopically named S($J_l$), where the $J_l$ is the lower level and the quantum rotational levels of the
transition are $J=(J_l+2) \rightarrow J_l$. Besides the pure rotational lines, transitions among
vibrational levels (quantum number $v$) are permitted (with $\Delta v=1$), 
leading to a wide range of ro-vibrational lines, such as 2--1 S(1), which is the $v=2\rightarrow 1$ and
$J=3\rightarrow 1$ transition.
Spectroscopic studies were focused on peaks within the $\omega$ ridge, primarily peaks B, C, and G.
Near-infrared spectroscopy detected ro-vibrational lines including 1--0 S(0) and S(1)  and 
2--1  S(0) through S(3) \citep{richter95a}. The populations of the upper levels are far from
any single-temperature equilibrium distribution, requiring either a range of temperatures or non-thermal 
excitation mechanism.
The pure rotational lines (i.e. vibrational quantum numbers $0\rightarrow 0$) are in the mid-infrared,
with the S(2) line detectable from the ground \citep{richter95a} and studied in more detail
beginning with the pioneering {\it Infrared Space Observatory}: 
\citet{cesarsky99} detected the S(2) through S(8) lines, and they found that the line brightnesses could not be well fit with a
single shock model but could be explained by a non-steady C-type shock into gas with preshock density $10^4$ \cm3 with shock velocity 30 \kms
that has only hit the dense gas within the last 2000 yr, so that
it has not yet reached steady state.
The spectrum between 2--5 $\mu$m was measured with {\it Akari} including H$_2$ lines with a range of excitation, 
leading to a conclusion that the 1-0 S(1) line is mainly from J shocks propagating into interclump media \citep{shinn11}.
A different conclusion was obtained from studying the {\it Spitzer} spectra, from which
\citet{neufeld07} concluded that the pure rotational lines with $J_l>2$ 
``likely originate in
molecular material subject to a slow, nondissociative shock that
is driven by the overpressure within the SNR.''



\def\tnm{\tablenotemark}
\begin{deluxetable*}{cccccccc}
\tablecolumns{8}
\tablecaption{Observability of H$_2$  Rotational Lines\label{h2lines}} 
\tablehead{
\colhead{} &\colhead{} & \colhead{} & \multicolumn{2}{c}{Brightness\tnm{b}} & \colhead{~} & \multicolumn{2}{c}{Atmospheric Transmission}\\
\cline{4-5}
\cline{7-8}
\colhead{Transition\tnm{a}} & \colhead{Wavelength} & \colhead{$E_{up}/k$} & \colhead{J-shock} & \colhead{C-shock} &
\colhead{} &  \colhead{14,000 ft} & \colhead{41,000 ft} \\ 
\colhead{} & \colhead{($\mu$m)} & \colhead{(K)} 
}
\startdata
S(0) & 28.21883 & 510 & $<0.02$ & 0.001 &&  0 & 0.8 \tnm{c} \\
S(1) & 17.03483  & 1015 & $<0.02$ & 0.04 && 0.45\tnm{c} & 0.98\\
S(2) & 12.27861  & 1682  & & 0.08 && 0.95 & 1 \\ 
S(3) & 9.66491   & 2504 & 0.14 &  0.7 && 0.2\tnm{c} & 0.2\tnm{c} \\
S(4) & 8.02505  & 3474  &  &   && 0.9\tnm{c} & 0.99\\
S(5) &  6.90952 & 4586 & 0.7 & 1 &&  0 & 0.98\\
S(6) &  6.10856 & 5829 &  &   && 0 & 0.9\tnm{c}\\
S(7) &  5.51116   & 7197 & 1 & 0.15 && 0.25\tnm{c} & 0.99\\
S(8) & 5.05303  & 8677 &  &  && 0.85\tnm{c} & 1\\
S(9) & 4.69461   & 10263 & 1 & $<0.001$ && 0.62\tnm{c} & 0.78\tnm{c}\\
1-0 S(1) & 2.1218 & 6956 & 0.2 & 0.1 && 1 & 1 \\
\enddata
\tablenotemark{a}{All transitions are pure rotational (0-0) except for the near-infrared 1-0 S(1) line.}
\tablenotetext{b}{Brightness of each line relative to the predicted brightness for an 80 \kms J-type shock into gas with density $n_0=10^3$ \cm3 \citep[see Fig. 5b of ][]{hm89}
or a 35 \kms C-type shock into gas with density $n_0=10^4$ \cm3 \citep[see Fig. 5b of ][]{drd83}. }
\tablenotetext{c}{Nearby deep atmospheric absorption line makes observation extremely challenging.}
\end{deluxetable*}

\section{New Observations with SOFIA}

The purpose of our new observations is to determine the velocities of the H$_2$ in the shocked clumps of IC~443.
To first order, the experiment is straightforward: if the shocks are fast (J-type), then 
motions of 50 \kms or greater are expected; on the other hand, if shocks are slow and magnetic-dominated (C-type),
then broad lines with width $\simlt 30$ \kms are expected. 
The distinction occurs at the shock velocity above which H$_2$ is dissociated. 
In C-type shocks, the H$_2$ is accelerated and should have motions spanning from the systemic velocity of the pre-shock cloud up to $\sim 30$ \kms in the direction of shock propagation.
In J-type shocks, the H$_2$ is dissociated but reforms as the gas cools behind the shock; the accelerated cooling
layer can have relative velocities up to the shock speed.
Velocity-resolved observations to date have primarily been of tracer species, in particular millimeter-wave heterodyne spectra of rotational transitions of the CO transitions, which are bright due to the high dipole moment of
the molecule and its reasonably high abundance. The CO line widths are typically 30 \kms
\citep{denoyer79,wangscoville92}. Spectroscopy of H$_2$O showed a wider range of velocities with one
position (clump B) showing emission spanning at least 70 \kms \citep{snell05}.
A velocity-resolved observation of the near-infrared H$_2$ 1-0 S(1) line using a Fabry-Perot showed line widths
typically 30 km/s, combination J and C \citep{rosado07}, with all these papers leading to combinations of J-type and C-type shocks to explain different aspects.

To advance understanding of the molecular shocks, we searched for suitable emission lines from the dominant species (H$_2$) that could be detected from a wide range of shock conditions and can be resolved sufficiently to measure speeds in the range 30--150 \kms. 
Table~\ref{h2lines} compiles the brightness conditions for two shock models relevant to the molecular shocks in IC~443.
The S(3) through S(9) lines are the brightest for one or the other type of shock, but only S(5) is bright in both types. 
These spectral lines are at mid-infrared wavelengths that are sometimes heavily absorbed by the Earth's atmosphere.
Table~\ref{h2lines} lists the atmospheric transmission at the wavelength for each line, for an observer at the specified altitude above sea level and $39^\circ$ latitude pointing toward a source with zenith angle $45^\circ$, 
calculated using ATRAN \citep{lord92}. The S(5) line cannot be detected from mountaintop observatories, with
essentially zero atmospheric transmission. But for a telescope at altitudes above the tropopause, the atmospheric transmission is 98\% and the observations are feasible.

We observed IC~443 using the Stratospheric Observatory for Infrared Astronomy \citep[SOFIA, ][]{young12,ennico18}, which has a 2.7-m primary mirror and operates at 39,000 to 43,000 ft (11.9 to 13.1 km) altitudes. At this altitude, SOFIA is generally above the tropopause when operating at non-tropical latitudes, and being in the stratosphere means SOFIA is above 99.9\% of 
water in the atmosphere. This affects the transmission of H$_2$ lines significantly; Table~\ref{h2lines} shows significant improvement except for the S(3) line, for which 
the  low transmission at both altitudes is primarily due to ozone, which has a much higher scale height than water.
Our observations focus on the H$_2$ S(5) pure rotational line at 6.90952 $\mu$m wavelength, or 1447.28 cm$^{-1}$ wavenumbers. 
We used SOFIA's mid-infrared Echelon-Cross-Echelle Spectrograph \citep[EXES, ][]{richter18}, which enables spectroscopy at wavelengths from 4.5-28.3 $\mu$m spanning the range of H$_2$ lines in Table~\ref{h2lines}. 
Practically, the S(1), S(2), S(4), S(5), S(6), S(7), and S(8) lines are likely to be feasible with Doppler shifts to avoid
nearby telluric absorption features. The S(2) and S(4) lines can be observed from the ground, but they are not predicted to be bright (nor is the S(1) line); the S(7) and S(8) lines are observable but not expected to be 
bright in slow shocks. This leads to S(5)  being the best suited line for studies of H$_2$ from molecular shocks, lacking observing capability from
space. For the near future,  the James Webb Space Telescope will have mid-infrared spectroscopy covering all these lines, but only at low spectral resolution $R<3000$ that cannot probe motions of gas in shock fronts in molecular clouds. Ground-based capabilities could make use of the S(2) line, 
e.g. with TEXES \citep{texesref}. The near-infrared allows access to a wide range of vibrationally excited lines with potential to bear on this problem, e.g. with IGRINS \citep{igrinsref}.

\def\tnm{\tablenotemark}
\begin{deluxetable*}{ccccccccc}
\tablecolumns{8}
\tablecaption{SOFIA Observing Log\label{obstab}} 
\tablehead{
\colhead{Date} & \colhead{Flight} & \colhead{Mode\tnm{a}} & \colhead{Altitude} & \colhead{Elevation} & \colhead{Time} & \colhead{PA$^b$} & \colhead{$V_{\rm cen}$} & \colhead{FWHM}\\
\colhead{} & \colhead{} & \colhead{} & \colhead{(feet)} & \colhead{($^\circ$)} & \colhead{(min)} & \colhead{($^\circ)$}
& \colhead{(km~s$^{-1}$)} & \colhead{(km~s$^{-1}$)}\\
}
\startdata
\sidehead{IC~443 C (06:17:42.4 +22:21:22): peak of 2MASS K$_s$ image in IC~443C}
2017 Mar 23 & 391  & Medium & 40,079 & 33 & 9 & 315 & -41.9 & 49.7\\
2019 Apr 19 & 555  & High   & 40,332 & 35 & 40 & 323 &-39.7& 41.0\\
\hline
\sidehead{IC~443 G (06:16:43.6 +22:32:37): peak of 2MASS K$_s$ image in IC~443G}
2017 Mar 21 & 389  & Medium  & 39,015 & 32 & 10 & 307 & -42.3 & 35.2\\
2019 Apr 24 & 558 &  High    & 39,005 & 33 & 17 & 323 & -38.5 & 17.4\\
\hline
\sidehead{IC~443 B1 (06:17:16.1 +22:25:15): peak of {\it Herschel} H$_2$O image in IC~443 B}
2018 Oct 18 & 514  & High      & 40,006 & 38 & 12 & 233 & -29 & 54.2\\
2019 Apr 26 & 560  & Medium & 39,011 & 28.1 & 15 & 327 & -38&53.9\\
\hline
\sidehead{IC~443 B2 (06:17:14.0 +22:23:16): peak of 2MASS K$_s$ image near IC~443 B}
2017 Mar 16 & 387  & Medium & 40,963  & 29 & 18 & 307 & -58.1 & 41.6 \\
2019 Apr 19 & 555  & High    & 39,806 & 45  &   & 323 & ... & ... \\
2019 Apr 26 & 560  & Medium & 39,029 &33 & 15 & 334 &-38 & 45.6 \\
\enddata
\tablenotetext{a}{Instrument mode: Medium=long-slit, medium spectral resolution ($R=13,000$). High=short-slit, high spectral resolution ($R=67,000$).}
\tablenotetext{b}{Position angle of slit, degrees E of N}
\end{deluxetable*}

Table~\ref{obstab} shows our SOFIA observation conditions. To resolve motions smaller than 10 \kms requires a spectral resolution
greater than $R\equiv\lambda/\Delta\lambda > 30,000$. We used EXES in its long-slit, medium-resolution $R=13,000$ mode to locate the peak emission for each clump and to search for gas moving faster than 30 \kms. Spectral lines were visible in near-real time with this mode.
To fully resolve the lines, we used the high-resolution mode, with $R=67,000$ capable of detecting motions of 4.5 \kms.
This is comparable to the turbulent motions in unshocked molecular clouds, so the emission from
any gas shocked by the blast wave should be spectrally resolved int he high-resolution mode.
The observing directions were chosen to be near well-studied molecular shocks, in order to add our new observations to the 
inventory of data being collected on these targets. The angular resolution possible with SOFIA/EXES is relatively high compared to
prior spectral observations. We set our initial positions 
using high-resolution narrow-band images of near-infrared H$_2$ 1-0 S(1)
and also an H$_2$O from Herschel for IC~443B1. The position IC~443B2 is away from the more commonly observed
spot in clump B; it was chosen based upon brightness of 2MASS K$_s$-band and unique color in the {\it WISE} images. 
The usable slit lengths are $9''$ and $72''$ for medium and high-resolution mode, respectively, and a slit width of $2.4''$ was
used in all observations.
The location and orientation of the medium-resolution slit for each source is
shown in Figure~\ref{wise443}.

Each spectrum was created by nodding the telescope between the clump position and a reference position, which was
chosen to be as close as possible but such that the 2MASS K$_s$ and WISE 4.5 $\mu$m images had no significant brightness when the long slit was placed at the reference position. The nod-subtracted spectra for all clumps showed emission at the
requested position, and clumps C and G also showed additional emission, resolved into multiple peaks, within $15''$ of the center.
The locations of the peaks correspond to the extent of the {\it Spitzer} 4.5 $\mu$m image (Fig.~\ref{wise443}) and spectral image
\citep{neufeld07}. 
We extracted the spectrum of the brightest peak, which was always near the center of the slit. To measure the medium-mode velocity resolution, we extracted an 
atmospheric spectrum
from the part of the slit that had no H$_2$ line from the target. The line widths
were 22.4 km~s$^{-1}$ (FWHM), for a resolving power $R=13,300$.

\begin{figure*}
\centering
    \plotone{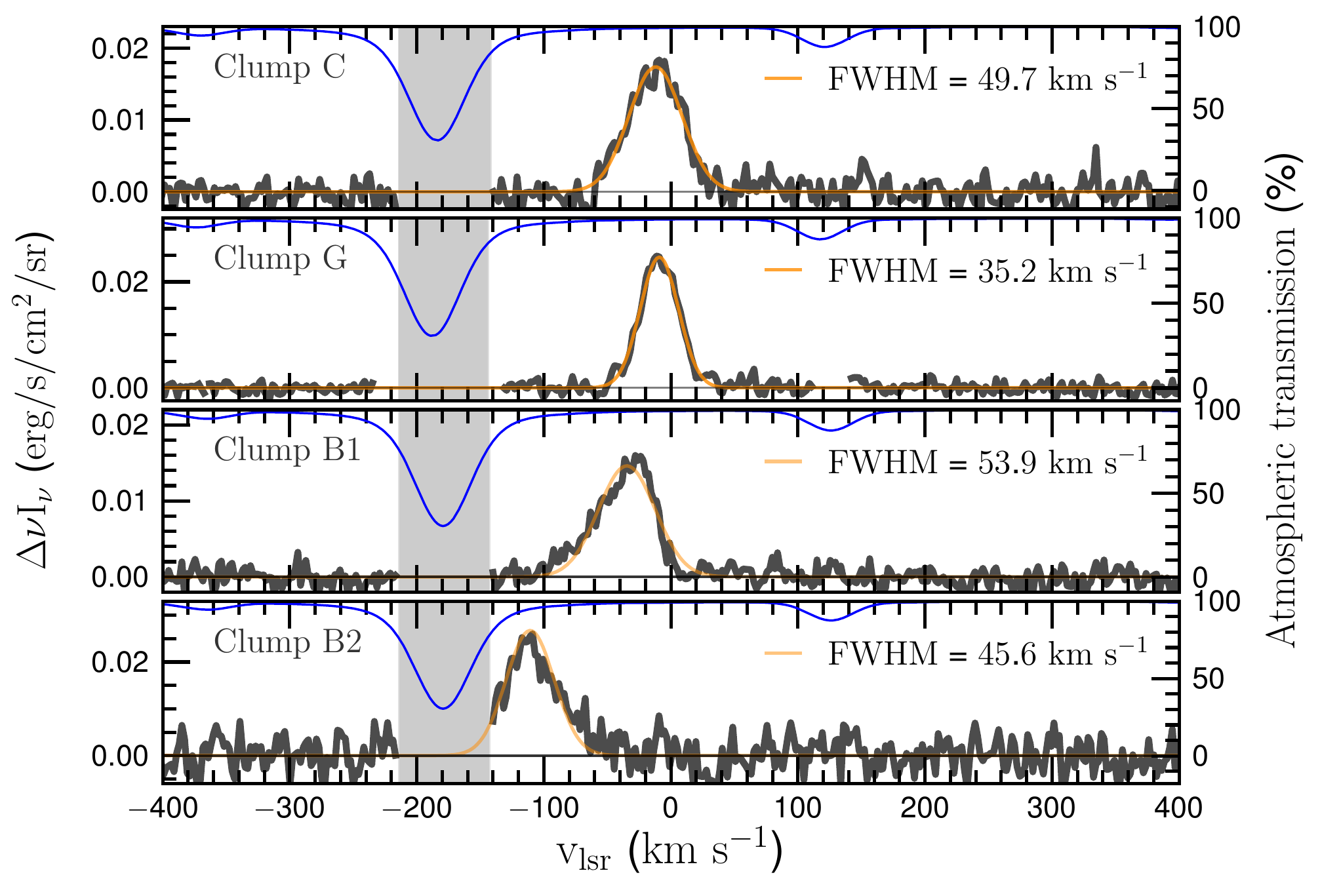}
    \caption{Medium-resolution H$_2$ S(5) 6.90952 $\mu$m ($\nu=1447.28$ cm$^{-1}$) spectra of (from top to bottom) clumps C, G, B1, and B2 in IC~443. In each panel, the background-subtracted intensity is shown in black. The intensity is in units of line brightness per spectral resolution element, $\Delta\nu I_\nu$, where $\Delta\nu=0.11$ cm$^{-1}$. The S(5) lines are marginally resolved at the resolution shown here. The atmospheric transmission is shown as a blue curve scaled to 100\% at the top of each panel. The grey region shows where the transmission is less than 50\%, which is where the background subtraction is less accurate. {\bfc The orange line shows a Gaussian fit with full width at half-maximum intensity (FWHM) labeled.}
\label{h2s5med}}
\end{figure*}

Figure~\ref{h2s5med} shows the medium-resolution spectra for each of the 5 clumps we observed in IC~443. Table~\ref{obstab} lists the central velocity and FWHM of a
simple Gaussian fit to each line. 
In all cases, the H$_2$ lines from IC~443 are
wider than the instrumental resolution. 
Also in all cases, the S(5) line was
separated from nearby atmospheric absorption features. 
The worst case of interference from atmospheric absorption is for IC~443~B2,
because the line from that source was
close to the deep absorption feature that has zero transmission from 6.9062 to 6.9064 $\mu$m (-144 to -135 km~s$^{-1}$ heliocentric
velocity for the S(5) line). 
Because the large velocity shift of IC~443~B2 relative to the ambient gas was unexpected, 
and because of the potential influence of
the atmospheric feature, we repeated the pilot observation (taken in 2017 Mar) on our final observing flight
(in 2019 Apr);
the results were nearly identical, confirming the original result. 

\begin{figure*}
\centering
    \plotone{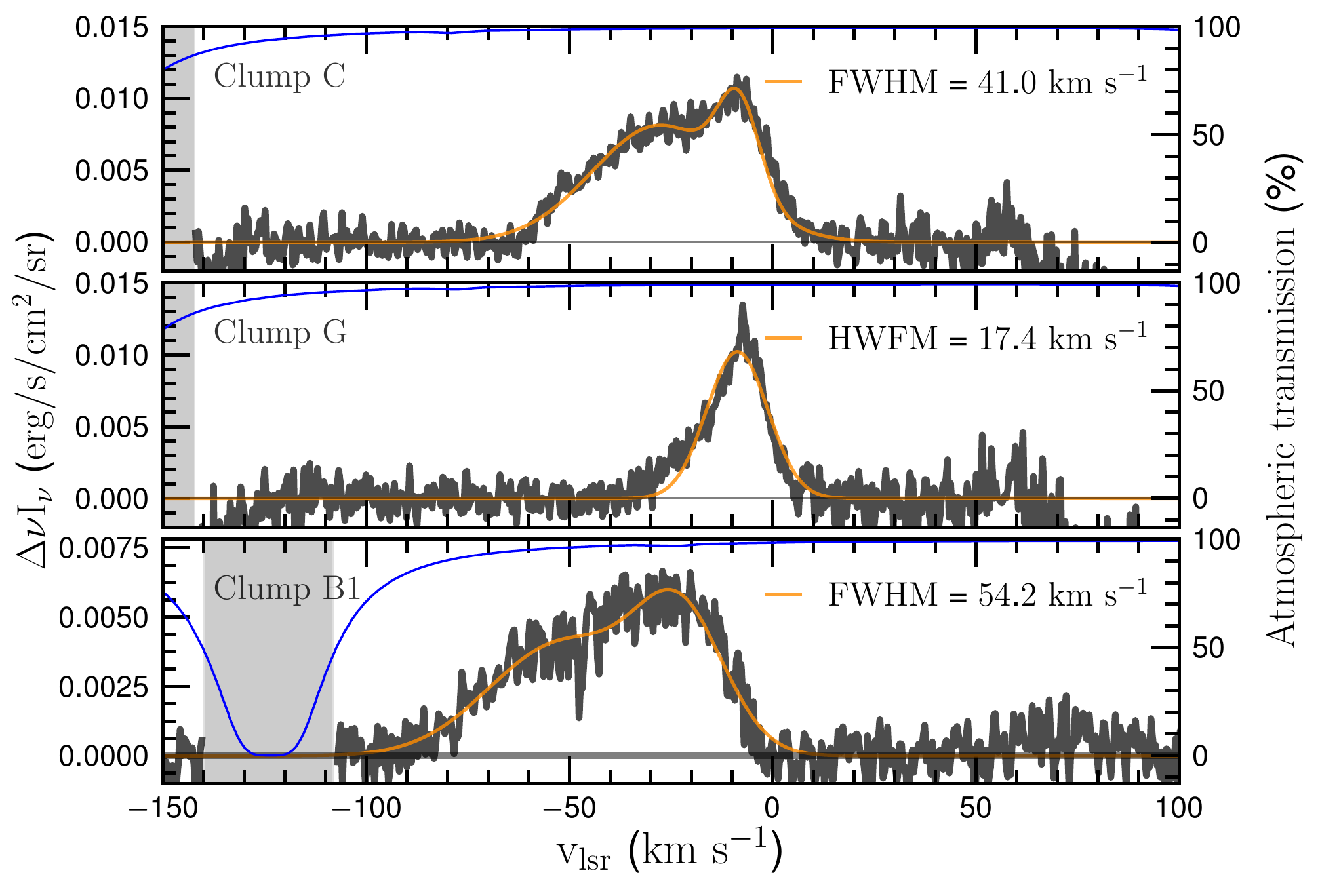}
    \caption{High-resolution H$_2$ S(5) 6.90952 $\mu$m ($\nu=1447.28$ cm$^{-1}$ spectra of (from top to bottom) clumps C, G, and B1 in IC~443. In each panel, the background-subtracted intensity is shown in black. The intensity is in units of line flux per resolution element, $\Delta\nu I_\nu$, where $\Delta\nu=0.022$ cm$^{-1}$. The S(5) lines are very well resolved at this resolution. The atmospheric transmission is shown as a blue curve scaled to 100\% at the top of each panel. The grey region shows where the transmission is less than 50\%, which is where the background subtraction is less accurate. The spectra are shown versus the
    velocity with respect to the local standard of rest (LSR) relative to nearby stars. The atmospheric absorption features appear at different velocities in each panel because they were observed on different dates, when
    the velocity of Earth relative to the LSR differed. 
    {\bfc The orange line shows a 2-Gaussian fit, with the FWHM of the wider component labeled.}
\label{h2s5high}}
\end{figure*}

Figure~\ref{h2s5high} shows the high-resolution spectra for the 3 IC~443 clumps that were detected at
high resolution. These spectra cover 
the same positions as those in Figure~\ref{h2s5med}. 
All positions were observed until a high signal-to-noise was obtained. 
For the position with the faintest emission in medium-resolution, IC~443~B2, the high-resolution spectrum 
did not
show a convincing detection. 
The line would have been detected at high resolution if it were narrower than  12 km~s$^{-1}$,
so we conclude that the line from IC~443~B2 is both broad and significantly shifted from the pre-shock velocity,
as determined from the medium-resolution observations. 

Finally, we estimate the amount of extinction by dust and its effect on the observed line brightness. 
One reason that shocks into dense gas have remained an observational challenge is that they occur,  by construction, in regions of high column density. Among the known supernova-molecular cloud interactions, IC~443 has relatively lower extinction, making it accessible to near-infrared (and, toward the NE, even optical) study. 
The extinction toward the shocks must be known in order to convert the measured brightness into emitted surface brightness, and to make an accurate comparison of ratios of lines observed at different wavelengths. For the prominent shocked clumps in IC~443, 
the estimated visible-light extinctions are 
$A_{\rm V}$=13.5 for Clump C and B and $A_{\rm V}$=10.8 for clump G \citep{richter95}. At the wavelength of the H$_2$ S(5) line we observed with SOFIA, the extinction is only 0.2 magnitudes, so the observed brightness is only 20\% lower than the emitted.


\section{Predictions of MHD shock models for IC~443 \label{sec:model}}

To understand the implications of our observations in terms of physical conditions in the shock fronts, we proceed
as follows. First, we assess the likely range of physical properties of the pre-shock gas and the shocks that
are driven into it. Then, we generate shock models {\it a priori} given those physical conditions.
Then in \S\ref{sec:compare}, we compare the shock models to the observations and discuss their implications.

\subsection{Conditions of the shocked and pre-shock material}

The properties of the gas before being impacted by the SN blast wave are of course not known; however, there is some direct evidence. \citet{lee12} measured the properties of nearby, small molecular clouds that are largely unshocked. The cloud they named SC05, for example, lies between shock clumps B and C; it was identified from CO(1--0) observations as a 13 $M_\odot$, cold clump with average density 
$\langle n_{{\rm H}_2}\rangle\sim 10^3$ cm$^{-3}$.
There also must exist higher-density cores within SC05 in order to explain emission from HCO$^+$, which has a critical density for excitation of $7\times 10^5$ cm$^{-3}$. 
In the interpretation by \citet{lee12}, SC05 is what
shock clumps B and C looked like, before they were crushed
by the SN blast wave. Cloud SC05 is at an LSR velocity of -6.2 \kms, which we will take as the nominal pre-shock 
velocity of the ambient gas. The CO(1--0) velocities for the ambient clouds around IC~443 have a range of
about $\pm 3$ km~s$^{-1}$ with respect to the average.

As mentioned in the Introduction, the pressure of the IC~443 blast wave appears inadequate to drive 
a blast wave that can  accelerate gas with the densities of a molecular cloud.
There are two keys to understanding the broad molecular lines.

{\it (1) Enhancement of pressure at the clump shock:}
\citet{moorhouse91} and \citet{chevalier99} showed that the part of the supernova remnant shell that impacts the clump is
caught between the impetus of the expanding shell and the inertia of the clump, and the pressure in the slab of
gas being driven into the clump is enhanced, relative to $p_{\rm ram}$, by a factor 
\begin{equation}
\frac{p_c}{p_{\rm ram}} = \frac{n_c}{n_0} \left[1 + \left(\frac{n_c}{n_{\rm shell}}\right)^\frac{1}{2}\right]^{-2},
\label{eq:popram}
\end{equation}
where $n_{\rm shell}$ is the density of the shocked gas of the shell being driven into the clump. 
We use the electron density of the shocked gas, 
based on optical spectroscopy of [S II] in the shocked  filaments \citep{fesen80}, 
to set $n_{\rm shell}=100$ \cm3. 
{\bfc Given the uncertainty in the preshock density for the overall supernova shell, we use a
 value $n_0=2$ \cm3\ that is within the range 0.25 to 15 \cm3\ of X-ray, optical, and radio constraints; this value corresponds to a supernova remnant age of 10,000 yr.}
The pressure enhancement is crucial, because otherwise, as explained above, the ram pressure of the blast wave is insufficient accelerate dense molecular gas to the observed velocities.

{\it (2) A range of densities for the clumps}: It is naturally expected that a molecular clump is not an instantaneous rise in density from that of the interclump medium to a single clump density. Rather, the dense clumps are more likely to be complicated structures with fractal or filamentary shape and density gradients. In this case, a wide range of shock velocities is expected. \citet{tauber94} considered the case of a centrally condensed clump with radial profile, and showed qualitatively that a 
bow-shock structure is expected, with the lower-density gas swept from the dense core away from the explosion center.
In addition to the structural implication of a range of densities, a much wider range of physical conditions is expected from a shocked clump than would arise from an isolated shock into a uniform clump of high density. 
The configuration would lead to emission along the same line of sight from gas with physical properties that cannot coexist
in the same location but rather arise from separate shocks into portions of the clump that have different densities.

Using equations 
\ref{eq:vs} to \ref{eq:popram}, the shock ram pressure and the pressure enhancement for shocks
into molecular clumps with a range of densities are listed in Table~\ref{shspeed}. 
We are not considering the topology of the cloud, where layers at different densities that are shocked first would alter the properties of the shocks into `downstream' portions of the cloud; instead, for this initial calculation, we imagine each
part of the cloud experiencing the direct blast wave simultaneously. 
Table~\ref{shspeed} lists the shock speed, $v_s$, into the gas at each density, ranging from 10 km~s$^{-1}$ in the highest-density gas up to 100 km~s$^{-1}$ in the lower-density, 
but still molecular, gas. The highest density into which a shock front will be driven is the one for which the pressure drives a Mach-1 sound wave into 
the clump, i.e. $v_s$ is greater than the sound speed. Using equations \ref{eq:pram} and \ref{eq:vs} for the shock pressure and speed, this condition becomes
\begin{equation}
n_c < 5\times 10^6 \left( \frac{T_c}{10\,{\rm K}}\right)^{-1} \left(\frac{p}{p_{\rm ram}}\right)\,{\rm cm}^{-3}. \label{eq:ncmax}
\end{equation}
At densities greater than $10^3$ cm$^{-3}$, the preshock material was no warmer than diffuse molecular clouds, which have $T_c\sim 100$ K, and likely they were as cold as molecular clouds, which have $T_c\sim 20$ K; therefore, shocks are driven into gas up to densities $n_c\sim 10^6$ cm$^{-3}$, which we take
as the upper density for our models.

To give an idea of the size of the shocked region and the relative locations of shock fronts into different densities,
Table~\ref{shspeed} then lists the distance that the shock front would travel, in $10^3$ yr, into gas of each density. 
The actual age of the shocks is not known, but should be less than the age of the SNR and more than the time since the shocks were first 
observed---that is, in the range 45 to 40,000 yr---so $10^3$ yr is a useful illustration that will
show the effects of early-time shock evolution.
We further list the separation of the various shock fronts on the sky, relative to the shock into 6,000 cm$^{-3}$ gas.
Relative to the shock into 6,000 cm$^{-3}$ gas, the 
shocks into higher-density gas (i.e., the molecular shocks)
will be in the `upstream' direction (negative separation), while shocks into lower gas (i.e., the ionic shocks) will have moved `downstream,' further away from the original SN explosion site. Table~\ref{shspeed} shows that only observations at the several-arcsec
angular resolution and several km~s$^{-1}$ spectral resolution
will distinguish the shocks with densities in the $10^3$--$10^5$
cm$^{-3}$ range. This is precisely where the SOFIA/EXES observations are unique. 

\begin{deluxetable}{ccccc}
\tablecolumns{5}
\tablecaption{Shock speeds into different densities for IC~443\label{shspeed}} 
\tablehead{
\colhead{$n_c$} &\colhead{$p/p_{\rm ram}$} & \colhead{$v_s$} & \colhead{$d$($10^3$ yr)} & \colhead{separation}\\
\colhead{(cm$^{-3}$)} & \colhead{} & \colhead{(km~s$^{-1}$)} & \colhead{(pc)} & \colhead{($''$)}}
\startdata
500     & 24 & 100 & 0.10  & 8\\
2,000   & 33 & 60  & 0.061 & 3\\
6,000   & 38 & 40  & 0.042 & 0\\
20,000  & 43 & 22  & 0.022 & -2\\
100,000 & 46 & 10  & 0.010 & -4
\enddata
\end{deluxetable}

\section{MHD shock model predictions}

\def\mhdvode{the Paris-Durham shock code}

We computed MHD shock models using \mhdvode{}, which incorporates a wide range of heating and cooling mechanisms and a full network of astrochemistry \citep{flowerpineau03,flower15,lesaffre13,godard19}.
Apart from using the specific conditions (pre-shock density and shock velocity) for IC~443, we used the following adaptations relative to the `nominal' parameters in the model.
First, we use an elevated cosmic-ray ionization rate,
$\zeta=2\times 10^{-15}$ s$^{-1}$,
based on the observations of H$_3^+$ \citep{indriolo10}.
The effect of photoionization on the shocked gas is included, because it was shown that irradiated shocks have different chemistry \citep{lesaffre13}.
The pre-shock magnetic field was set according to 
$B\simeq 1 n_{\rm H}^{0.5}$ $\mu$G, where $n_{\rm H}$ is the pre-shock density,
$n_{\rm H}=n_{\rm H{\small I}} + 2 n_{{\rm H}_2}$.
an approximate match to observations of molecular clouds 
with densities in the $10^3$ to $10^5$ cm$^{-3}$ range \citep{crutcher12}.
Because the SN explosion was relatively recent, the time required to establish a steady-state shock front is, for some combinations of 
density and shock speed, longer than the age of the supernova remnant. We therefore consider non-steady shock models that evolve 
only up to a time $t$, and we ran models for $t=10^3$ and $10^4$ yr. The age of the SN explosion has been estimated in the range $4\times 10^3$ to
$3\times 10^4$ yr \citep{troja08,chevalier99}, and it took some time for the shock front to propagate to the distance from center where we
observe the shock clumps now. The $10^3$ yr models would apply if the blast wave reached the clumps at 75\% of the 
lower-estimated 
age of the SN explosion and 96\% (i.e. very recently) of the upper-estimated age of the SN explosion.
The $10^4$ yr models are older than the lower-estimated age of the SN explosion (so they cannot apply if that age is correct) and
67\% of the upper-estimated age of the SN explosion.
the SN explosion.

\begin{deluxetable*}{ccccccc}
\tablecolumns{7}
\tablecaption{Properties of model shocks for different densities\label{shmod}} 
\tablehead{
\colhead{$n_c$} & \colhead{$v_s$} & \colhead{$\langle z\rangle$} & \colhead{$I({\rm S}5)$} & \colhead{FWHM} & 
\colhead{$t_{\rm steady}$} &
\colhead{Shock Type}\\
\colhead{(cm$^{-3}$)} & \colhead{(km~s$^{-1}$)} & \colhead{(pc)} & \colhead{($10^{-3}$erg~s~cm$^{-2}$s~r$^{-1}$)} &
\colhead{(km~s$^{-1}$)} & 
\colhead{(yr)} }

\startdata
500   & 100 & $\sim 0.1$ & 0.02 & $\sim 1$ & 200,000 & J \\
2,000 & 60  & 0.016      & 1.5  & 28 & 50,000 & CJ \\
6,000 & 37 & 0.011      & 8.9  & 19 & 20,000 & CJ \\
20,000 & 22 & 0.0018     & 9.8  & 10 & 5,000 & CJ \\
100,000 & 10 & 0.0003     & 7.7  & 4  & 1,000 & C 
\enddata
\end{deluxetable*}

The shocks into the gas have different character depending on their physical conditions. A key condition is whether the shock can be treated
as a single fluid (in which case all material moves at the speed of the primary mass contributor, which is H$_2$) or the ions and neutrals
separate (in which case a 2-fluid, C-type shock develops).
The distinction occurs for shocks with speed lower than the magnetosonic speed,
\begin{equation}
    v_m^2 \equiv c_s^2 + v_{Ac}^2
\end{equation}
where $v_{Ac}$ is the effective Alfv\'en speed \citep{lesaffre13}.
The dynamics of the shock depend critically on the mass density of ionized particles that couples with the magnetic field.
The gas in molecular clouds is ionized by cosmic rays to an ionization fraction $x=n_i/n_{\rm H}\simlt 10^{-7}$.
With the gas predominantly neutral, charged dust grains become the dominant mass loading onto the magnetic fields \citep{guillet07}.
The mass-weighted average absolute value of the grain charge, $\langle|Z|\rangle$, is of order unity under conditions of
low radiation field \citep[cf.][]{drainesutin87,ibanez19}. 
We write the magnetic field in terms of the gas density as, $B = b (n_{\rm H}/{\rm cm}^{-3})^{1/2}$~$\mu$G, a scaling that applies to flux-frozen magnetic fields,
with observations showing average magnetic fields consistent with $b$ of order unity \citep{crutcher12}.
The criterion for a C-type shock, $v_s<v_m$, then can be expressed as a limit on the magnetic field strength:
\begin{equation}
    b > \frac{v_s}{19\,{\rm km~s}^{-1}} \left[ 100 x + \langle|Z|\rangle  
    \frac{100}{[{\rm G}/{\rm D}]}
    \right]^{\frac{1}{2}}
    \left(1-\frac{1}{M^2}\right)^\frac{1}{2}, \label{eq:smallb}
\end{equation}
where $[{\rm G}/{\rm D}]$ is the gas-to-dust mass ratio, and $M=v_s/c_s$ is the Mach number (which is greater than 1 as enforced by equation~\ref{eq:ncmax}).
The first term in the square brackets is for ionized gas; in molecular clouds, $x<10^{-6}$ so this term is small.
The second term in the square brackets is for charged grains and is of order unity; this
term will dominate unless the
gas is highly ionized or the grains are uncharged, neither of which are likely to apply.

Referring to the preshock conditions from Table~\ref{shspeed}, we see that two-fluid shocks occur (i.e. the
inequality in equation~\ref{eq:smallb} applies) for magnetic field strength 
$B>100$ $\mu$G for $n_c=10^3$ cm$^{-3}$ and 
$B>350$ $\mu$G for $n_c=10^5$ cm$^{-3}$,
comparable to the observed average fields in molecular clouds.
If we consider the environment around the IC~443 progenitor as having been similar to the Orion Molecular Cloud, 
average magnetic field strengths of 300 $\mu$G were derived from dust polarization \citep{chuss19}. 
It is thus expected that the conditions for C-type shocks are present, which is important for generating a significant layer of dense, warm, molecular gas that generates wide H$_2$ lines. 
The magnetic field in the shocked molecular gas was directly detected via the Zeeman effect for
radio-frequency OH masers, yielding strengths of 200 to 2200 $\mu$G \citep{brogan00,brogan13}.
For the shocks model calculations in this paper, we set the magnetic field input parameter $b=3$, which
allows for C-type shocks to develop. 

The results of the shock models for the input parameters from Table~\ref{shspeed}
relevant to IC~443 are as follows.
First, the shocks are single-fluid J-type for densities lower than $10^3$ cm$^{-3}$,
because the shock speeds are greater than the magnetosonic speed, unless the magnetic field
is significantly enhanced relative to a nominal molecular cloud. 
Second, the shocks for moderate-density ($10^3$ to $10^4$ cm$^{-3}$) gas are
non-steady, because they do not have time to develop to steady state within $10^3$ yr.
Third, shocks into high-density ($\simgt 10^5$ cm$^{-3}$) gas are steady-state because
cooling is so fast that they develop within $10^3$ yr.
Table~\ref{shmod} lists basic model properties including the spatial extent of the 
shock front, the total brightness of the S5 line of H$_2$, and
the time it takes for a steady-state shock to develop
\citep{lesaffre04}.
The shock fronts are generally thin; at the distance of IC~443, the shocks into $10^3$ cm$^{-3}$ clumps would
subtend, if viewed edge-on, $2.2''$, which would be nearly resolved in the {\it Spitzer}/IRAC observations and
resolved to SOFIA/EXES and high-angular-resolution near-infrared images \citep{richter95}, while the shocks into denser gas would be unresolved. 

\begin{figure}
    \centering
    \plotone{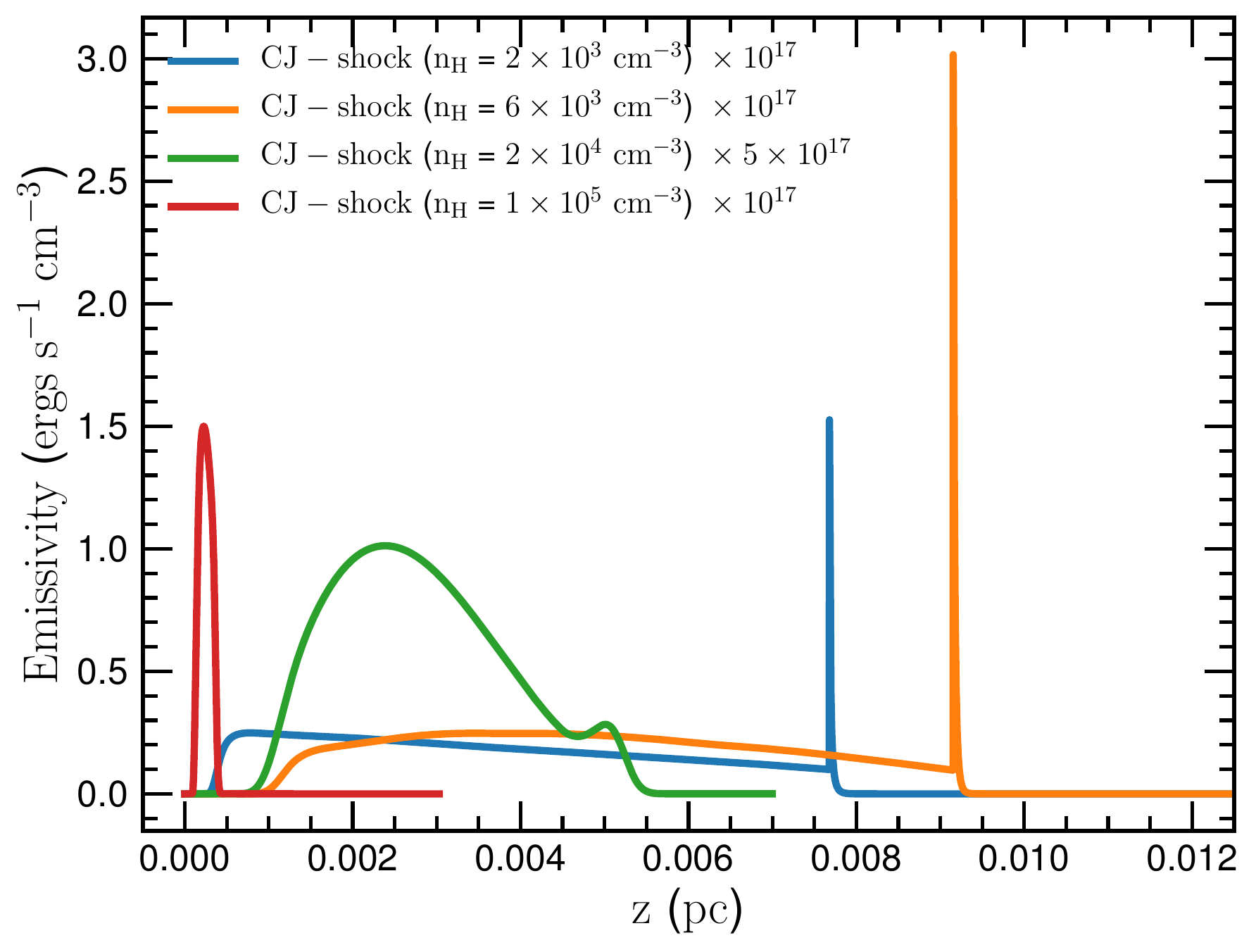}
    \caption{Emissivity of the H$_2$ S(5) line behind shocks with properties listed in Table~\ref{shspeed}. 
    The shock into the highest-density, $10^5$ cm$^{-3}$, gas is in a compressed region due to rapid cooling of the gas. Shocks into gas with density greater than $10^5$ down to $10^3$ cm$^{-3}$ are predominantly
    C-type, and the H$_2$ emission arises from a broad region whose size increases for decreasing density. 
    The shocks into gas with
    density $\le 10^4$ cm$^{-3}$ show an additional, spatially-narrow spike at the greatest distance ($z$) behind the shock
    front; this is the J-type shock that is present because the shock age ($10^3$ yr) is shorter than the time required to
    fully develop a steady-state C-type shock. }
    \label{fig:shthick}
\end{figure}

\begin{figure}
    \centering
    \plotone{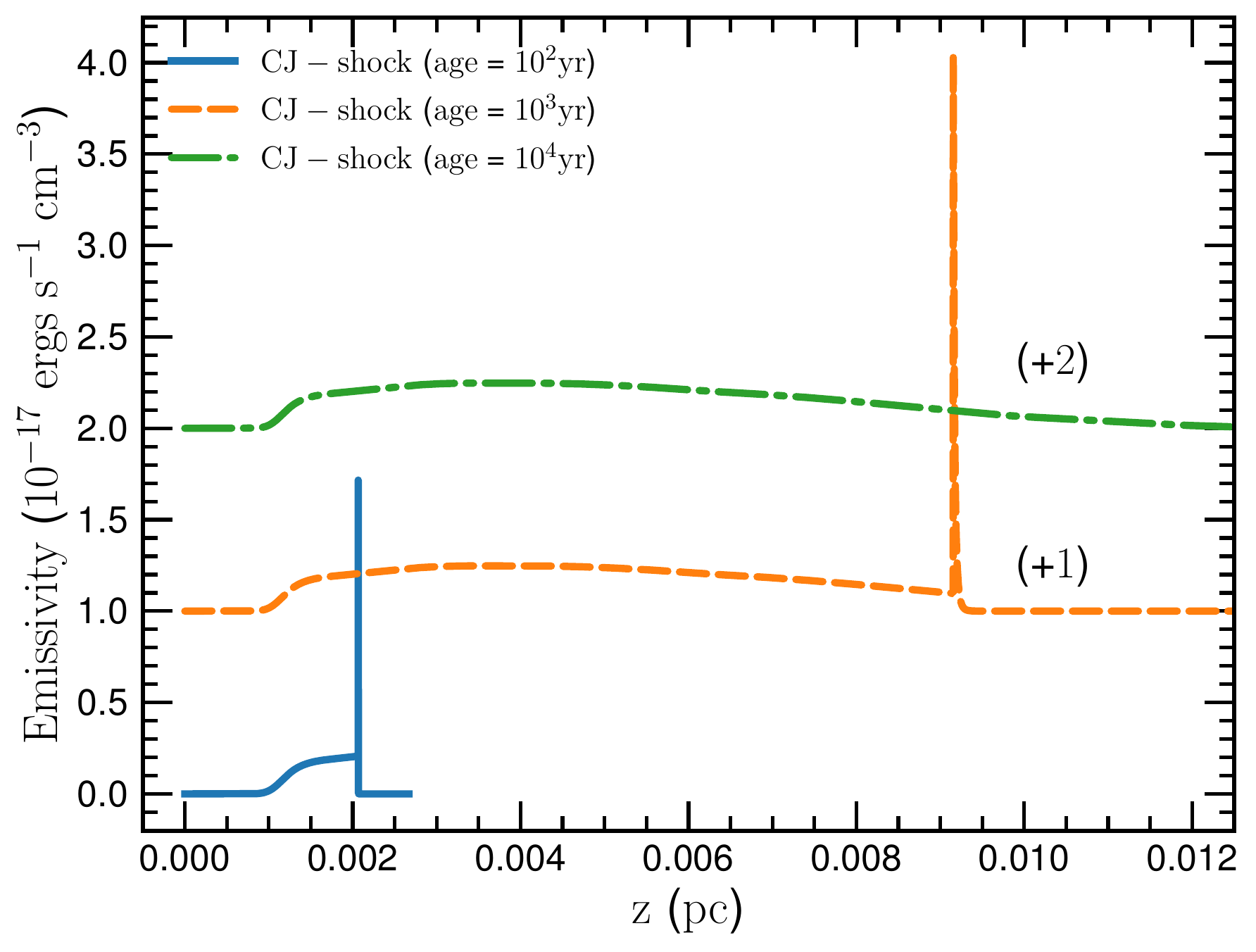}
    \caption{Emissivity of the H$_2$ S(5) line behind shocks into gas with preshock density $6,000$ cm$^{-3}$ and shock velocity 37 km~s$^{-1}$ with ages 
    $10^2$, $10^3$, and $10^4$ yr. The upper curve shows the oldest model, which approaches the shape of a fully-developed, steady-state C-type shock. 
    The lowest curve shows the youngest model, which has the distinct CJ-type shape: 
    the broad C-type region is truncated, and the J-type portion is relatively prominent. For the expected age of IC~443 shocks, shown in the middle curve, the shock is almost fully developed into C-type but with a small J-type component.}
    \label{fig:shtime}
\end{figure}

In order to calculate and understand the predicted line profile, we use the local emissivity 
$\mathcal{E}(z)$
versus distance
behind the shock, $z$.
Figure~\ref{fig:shthick} shows the emissivity versus distance for shocks with properties listed
in Table~\ref{shspeed}. Key characteristics of the shock models include the following.
(1) There is an offset of the emitting
region behind the shock front, due to the delayed reaction of the ions and neutrals; this offset increases
with decreasing preshock density (for fixed shock ram pressure). 
(2) The width of the emitting region increases with decreasing preshock density.
(3) For densities $10^3$ to $10^4$ cm$^{-3}$, the distinct characteristics of a CJ-type shock appear as a
spike in emissivity (due to high temperatures) far behind the shock front. This spike is the J-type component
of the CJ shock and appears because we only allowed the shock to evolve for $10^3$ yr, based upon 
arguments above that this is the likely age of shocks given the age of the supernova explosion. 
To clearly show the effect of shock age, Figure~\ref{fig:shtime} shows the emissivity versus distance 
for the preshock density $10^4$ cm$^{-3}$ case at three different ages.
For the  IC~443 blast wave, the CJ characteristics are expected to be mild
for shocks into $10^4$ cm$^{-3}$ and more prominent for $10^3$ cm$^{-3}$ gas.
(4) Finally, for the lowest density we considered, $10^2$ cm$^{-3}$, 
the temperature behind the shock is so high that 
molecules are dissociated and atoms ionized, 
leading to a J-type shock with an extended recombination zone. 
The H$_2$ emission only arises far behind the shock in the region after the gas has recombined and cooled 
to $\sim 300$ K
and compressed to where molecules can reform, at $z\sim 0.1$ pc \citep{hm89}, and the total emission from that shock is two orders of magnitude fainter (Table~\ref{shmod}) than for the C (and CJ) shocks.
Note that the reformation of H$_2$, behind J-type shocks into gas with densities less than 
$10^3$ cm$^{-3}$, takes $10^6$ yr, so a steady-state J shock of this sort cannot exist for a
supernova remnant like IC~443. Indeed, molecule reformation would only be complete for gas
with density $>10^5$ cm$^{-3}$, and shocks into gas with such high density are affected by the
associated strong magnetic field so that they do not destroy H$_2$.

\begin{figure}
    \centering
    \plotone{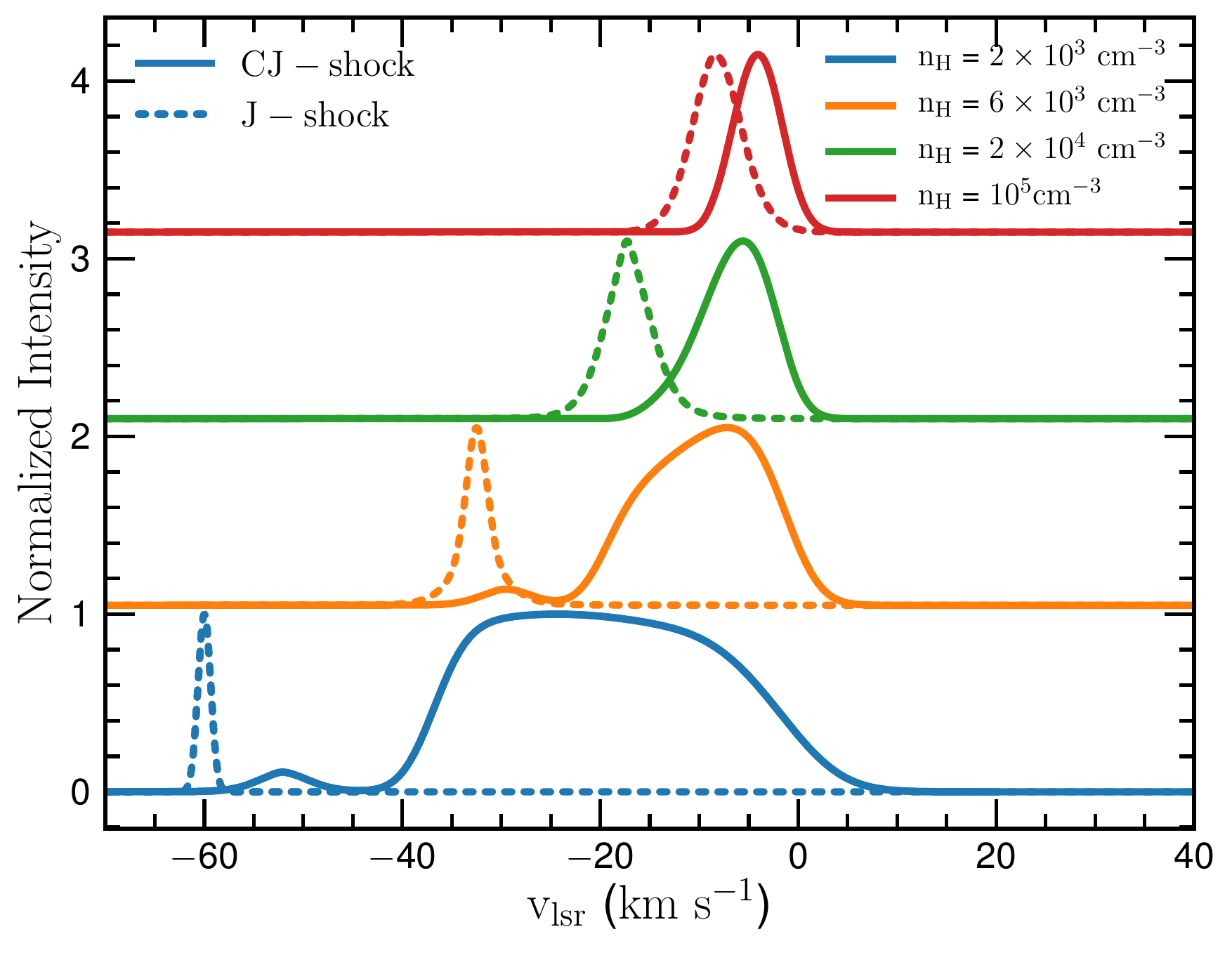}
    \caption{Velocity profiles of the H$_2$ S(5) lines produced by shocks with properties listed in Table~\ref{shspeed}. The models are for face-on shocks, moving toward the observer, and the horizontal axis is the projected (`observed') velocity in \kms\ {\bfc relative to the pre-shock gas at velocity zero.}
    The velocities with significant emission depend on the
    speed and temperature of the gas at each distance behind the shock (Fig.~\ref{fig:shthick}).
    The velocity profile for the highest-density shock is the most narrow and close to the zero velocity, due to rapid cooling of the dense gas. With decreasing density the shock becomes wider (because the same pressure drives faster shocks that heat the gas to higher temperature) and shifted to higher velocity. For the CJ-shock into $n_c=2,000$ and $6,000$ cm$^{-3}$ gas,  small peaks near the {\bfc shock} velocity are the emission from the non-steady portion of
    the shock as was illustrated in Figure~\ref{fig:shthick}.        }
    \label{fig:shmod}
\end{figure}

\begin{figure}
    \centering
    \plotone{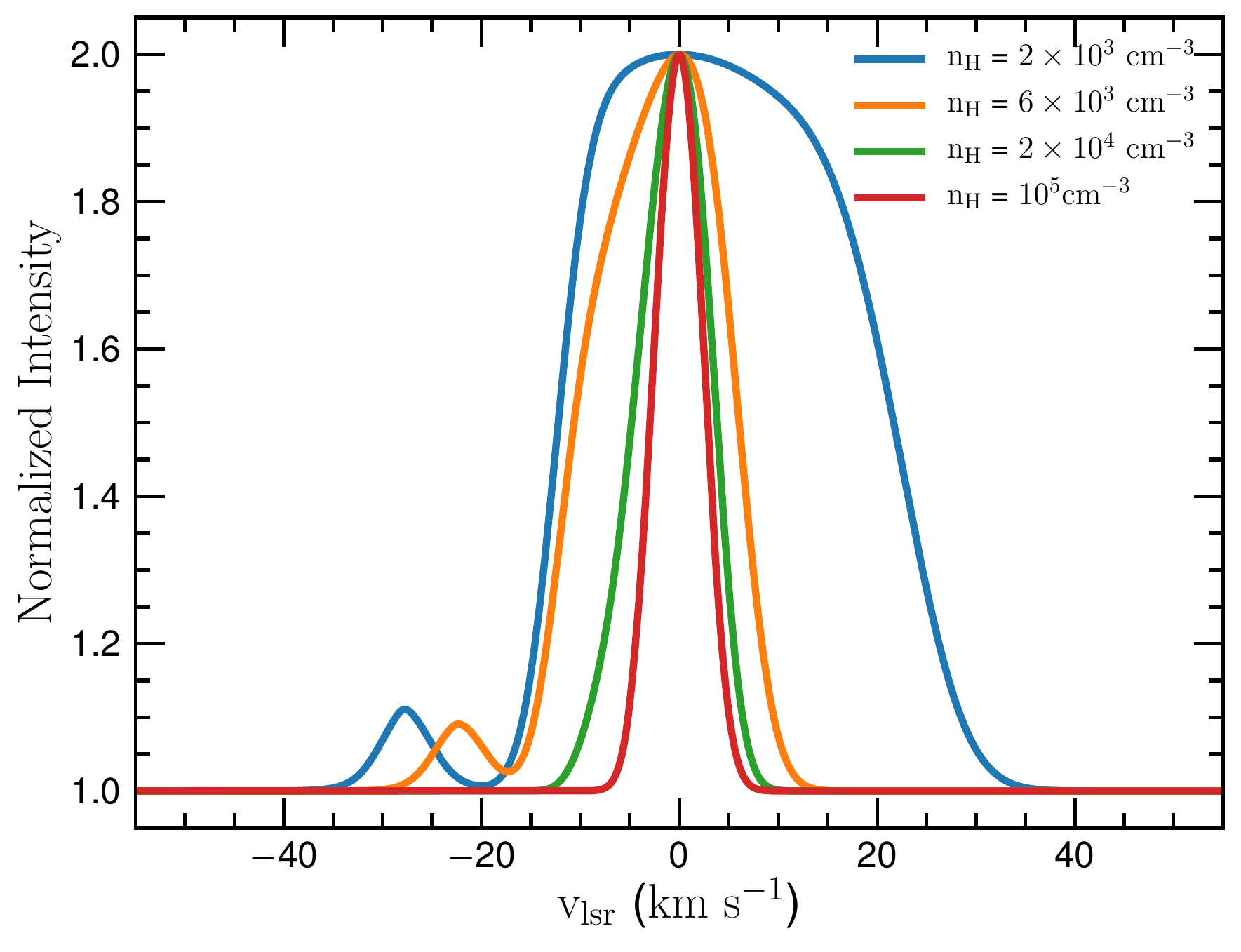}
    \caption{The same models as shown in Figure~\ref{fig:shmod}, shifted in velocity so that their peak is at 0 km~s$^{-1}$.
    The most narrow line profile is for the shock into the densest gas, with a steady progression of increasing H$_2$ line
    width as pre-shock density decreases. The detailed shape of the line profiles relates to the post-shock
    temperature and density structure. 
    For the shocks into $n_c=2,000$ and $6,000$ cm$^{-3}$ gas,  small peaks in the predicted line profiles due appears near the shock velocity due to hot gas just
    behind the shock front; this gas is still noticeably far from its steady state. If allowed to evolve
    for several thousand years, the line profiles would  be as wide on the
    negative-velocity side as it is on the positive-velocity side.
    The shifted profiles shown in this Figure are only for illustration of the line shape differences; for fitting observations, only the {\it unshifted} profiles in Figure~\ref{fig:shmod} should be used.}
    \label{fig:shmodshift}
\end{figure}

With the emissivities, we can now calculate the velocity profiles of
the H$_2$ S(5) lines by integrating over the line of sight:
\begin{equation}
    I(v) = \int {\rm d}z  \frac{\mathcal{E}(z)}{\sqrt{2\pi}\sigma(s)} 
    e^{-\left[v-\left(v_n(z)-v_s\right){\rm cos}i\right]^2/2\sigma(z)^2}
\end{equation}
where $i$ is the inclination of the shock front with respect to the line of sight, and $\sigma(z)$ is the thermal 
dispersion for gas at the temperature of neutral gas at distance $z$ behind the shock front.
The spectrum of the H$_2$ emission lines contains key diagnostic details for the properties of the shock fronts. 
Figure~\ref{fig:shmod} shows the H$_2$ S(5) line profiles from model shocks, and Table~\ref{shmod} lists the widths of the model-predicted lines. The predictions are for face-on shocks. 
For a shock front inclined relative to the line of sight, the line profile changes in a somewhat complicated manner that depends upon the emissivity and velocity at each
distance behind the shock. 
However, to first order, the width of the line profile is not strongly affected by inclination of the shock relative
to the line of sight, 
because the emission from each parcel of 
post-shock gas is isotropic, with velocity dispersion depending on its temperature.

For the lowest density in Table~\ref{shspeed}, $10^2$ cm$^{-3}$, the emission from the
J-type shock can be roughly approximated by a Gaussian with very narrow 
($\sim 1$ km~s$^{-1}$) velocity dispersion corresponding to the $\sim 300$ K temperature of the molecule reformation region \citep{hm89}. 
{\bfc  The J-shock front is at velocity  $v_0 + v_s$, where
$v_0$ is the velocity of the ambient, pre-shock gas.
The molecule reformation region where the gas has slowed 
would have velocity $v_0 + v_s/4$ (for a 4:1 density increase 
and conserving momentum),
so it would appear at approximately
-30 km~s$^{-1}$.} The amount of H$_2$ reformed behind such a shock is expected to be very small,
because the age of the supernova remnant is only 1\% of the time for the molecules to form.


\section{Comparison of Shock Models to Observations \label{sec:compare}}

First, we compare the predicted brightness of the H$_2$ S(5) line to that observed for the shock
clumps in IC~443.
The well-calibrated {\it Spitzer} spectrum of IC~443C  has $I({\rm S}5)\simeq 2\times 10^{-3}$ erg~s$^{-1}$~cm$^{-2}$~sr$^{-1}$
\citep{neufeld07}, and our SOFIA/EXES spectrum (at higher angular and spectral resolution) 
has $I({\rm S}5)\simeq 4\times 10^{-2}$ erg~s$^{-1}$~cm$^{-2}$~sr$^{-1}$.
The models with $n_c>10^3$ cm$^{-3}$ can explain the observed brightness, with modest secant boosts
to account for inclination of the shock front with respect to the line of sight.
For comparison to older shock models, the prediction from \citet{drd83} for shocks into gas with $n=10^4$ cm$^{-3}$ and
$v_s=37$ km~s$^{-1}$ agree well with \mhdvode. 
The J-shock models from \citet{hm89} are significantly
different: for $n=10^3$ cm$^{-3}$ and
$v_s=60$ km~s$^{-1}$, they predict a brightness $10^3$ times lower than the
observed lines; the assumptions in those J-type models and the time required to reach a steady state 
lead to predicted brightness far too low to explain the observed lines.


Second, we compare the velocity width of the lines to the model predictions. The observed widths are in the range
17 to 60 km~s$^{-1}$ FWHM (with non-Gaussian lineshapes extending to higher velocities. 
The model 
predictions for shocks into gas with density in the range $10^3$ to $10^5$ \cm3 (Table~\ref{shmod})
can explain most of this range. The narrow lines from reformed molecules in a 300~K layer behind J-type shocks 
\citep{hm89} cannot produce these lines because they are too narrow.
Similarly, shocks into higher-density gas may be present but cannot contribute much of the observed lines.
This conclusion is in accord with that derived in the previous paragraph based upon the line brightness.

Now, we can use the new {\it a priori} theoretical predictions for the detailed shape of shock line profiles to directly
compare to the observed spectra. We assume the pre-shock gas is within $\pm 3$ \kms\ of -6.2
km~s$^{-1}$ based upon the small, unshocked clouds near IC~443 \citep{lee12}.
The models are computed for `face-on' shocks that are infinite and plane-parallel. 
The observed brightness can be enhanced relative to the face-on value by the secant of the tilt, $i$,
between the line of sight and the shock normal: $I_{\rm obs}=I_0 {\rm sec}(i)$.
In practice, the
pre-shock gas is likely to be structured with a range of densities and finite physical scales,
{\it and} the shocks have non-trivial thickness. Thus the effect of the inclination and the beam
filling factor partially cancel each other out. If the shock thickness is $z$ (see Figure~\ref{fig:shthick}),
and the pre-shock clump size in the plane of the sky is $L$, IC~443 is at distance $D$, and the beam size is $\theta_b$, then, for an edge on shock, the beam filling factor is $z/D\theta_b$ and the boost in
brightness due to inclination is $L/z$. The product of beam filling factor and secant boost
is $L/D\theta_b$, which is the maximum 
boost in brightness relative to a face-on shock.
The 2MASS, {\it Spitzer}, and {\it Herschel} images show structure at the $L/D\simlt 2''$ scale, and the slit
size for SOFIA spectra is $2.4''$, so $L/D\theta_b\simlt 0.8$ is of order unity. The net effect of the
inclination effects on the shock brightness, relative to a face-on shock, is that it can be lower by 
a factor of $\sim 0.8$ or bright by a factor of ${\rm sec}(i)$ up to a factor of $L/D\theta_b \simlt 10$.
Thus there is very limited flexibility in using a shock model to explain the observed brightness:
a successful shock model must be between 80\% and 1000\%
of the observed brightness, in {\it absolute} units.

\begin{deluxetable*}{ccccccc}
\tablecolumns{5}
\tablecaption{Fits of shock models to H$_2$ S(5) from IC~443 clumps\label{modfit}} 
\tablehead{
\colhead{Clump} & \colhead{$n_H$} & \colhead{$v_s$} & \colhead{shock age} &\colhead{{\bfc Viewing angle}} & \colhead{{\bfc Scaling Factor}}  &\colhead{shock type}\\
\colhead{} & \colhead{(cm$^{-3}$)} & \colhead{(km~s$^{-1}$)} & \colhead{(yr)} & \colhead{($^o$)} &\colhead{} }
\startdata
C      & {\bfc 2,000}   & 60 & 5,000 & {\bfc 0} & {\bfc 1.07}  & CJ\\
       & {\bfc 10,000}  & 31 & 3,000 & {\bfc 0} & {\bfc 0.3} & CJ\\
G      & {\bfc 6,000}   & 37 & 3,000 & {\bfc 65} & {\bfc 0.3} & CJ\\
       & {\bfc 2,000}   & 60 & 5,000 & {\bfc 60} & {\bfc 0.33} & CJ\\
B1     & {\bfc 2,000}   & 60 & 5,000 & {\bfc 0} & {\bfc 0.14} & CJ\\
       & {\bfc 2,000}   & 60 & //    & // &//& J\\
B2     & {\bfc 400}     & 110& //    & // & // & J\\
\enddata
\end{deluxetable*}

\begin{figure*}
    \centering
    \plotone{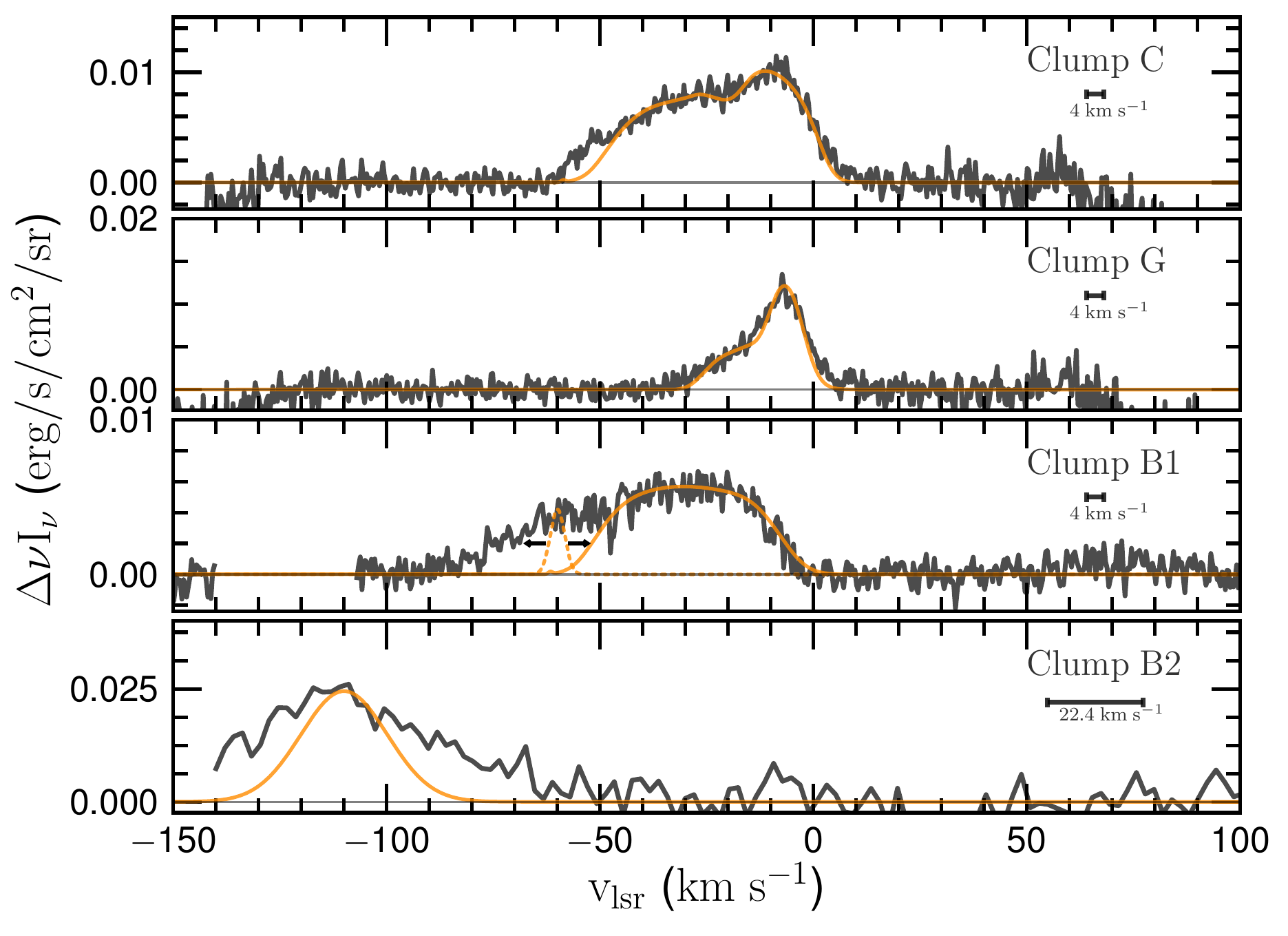}
    \caption{Shock model fits to H$_2$ S(5) emission from clumps C, G, B1, and B2 (top to bottom) in IC~443. The model components are listed in Table~\ref{modfit}. The spectral resolution is indicated by a scale bar to the right; for clump B2, only the `medium' resolution was observed. For clumps C, G, and B1, broad emission at velocities -50 to 0 km~s$^{-1}$ is due to CJ shocks with dynamics strongly affected by the magnetic field. 
    {\bfc The orange line shows the model fits described in the text.}
    For clump B2, there is no significant emission in that velocity range, and model is a J-type shock (with no magnetic field). For clump B1, we show a fit including CJ shocks for the -50 to 0 km~s$^{-1}$ range together with a single J-type shock for illustration; the emission from -90 to -50 km~s$^{-1}$ can be explained by a sum of such J-type shocks into a range of densities (indicated by the arrows next to the J0shock model, pointing toward lower and higher velocity).
    }
    \label{fig:fit_obs}
\end{figure*}

We can improve the agreement between the models and the observations by allowing freedom in
inclination of
the shock front with respect to the line of sight, by combining
shock models representing shocks into different-density gas, and by adjusting the pre-shock gas properties.
We restrict parameter tuning so that only models that have properties that are within reasonable 
expectation for IC~443 are considered. We only permit a $\pm 3$ km~s$^{-1}$ shift in the ambient gas velocity, which proves to be a powerful constraint now that we have a full theoretical model for the velocity profile
of the shocked gas. 
We require the shock age to be less than the age of the supernova remnant, estimated as 10,000 yr. 
Compared to the {\it a priori} models in \S\ref{sec:model}, which fixed the age at 1,000 yr, allowing for older shocks allows
the non-steady solutions for shocks into $10^3$ and $10^4$ cm$^{-3}$ gas to evolve toward full, steady-state
solutions. This more-complete evolution increases the width of the emergent H$_2$ line (and removes the
J-type contribution at the ambient velocity). 

The model fits to the observed H$_2$ lines are shown in Figure~\ref{fig:fit_obs}, with parameters summarized in Table~\ref{modfit}. 
The `scaling factor' in this table can be attributed to the boosts and decrements due to inclination of the shock and finite extent
of the clump as discussed earlier.
The emission observed at velocities -50 to 0 km~s$^{-1}$ can be explained as CJ-type shocks into gas with pre-shock densities in the range $10^3$ to $2\times 10^4$ cm$^{-3}$.
range of those observed for the pre-shock gas; the pressure of the shock fronts is specified from the {\it a priori} estimates based on X-ray and optical properties of
the supernova remnant. For each model, a shock age was derived by fit. The shocks into the highest-density gas could actually
be younger, because they rapidly reach their steady state.
But the shocks into the $10^3$ cm$^{-3}$ gas show ages of 5,000 yr. 
This age is older than some estimates of the entire age of the supernova remnant.
Allowing some time for the blast wave to reach the preshock cloud, our results support the older estimates that the supernova explosion occurred more than 6,000 yr ago.

The CJ-type shocks are all for magnetic fields oriented perpendicular to the shock: $\vec{B} \bot \vec{v_s}$.
For clump B2 and for the most-negative velocity emission from clumps B1 and C, emission from J-type shocks is
required. The critical difference, in terms of observable properties, is that when the dynamics of the shocked gas
are dominated by the magnetic field, the gas is gently and steadily accelerated, leading to a broad line profile that
spans from the velocity of the preshock gas up to a significant fraction of the shock velocity. On the other hand,
if the magnetic field is oriented so that it does not affect the shock dynamics, i.e. $\vec{B} \parallel \vec{v_s}$, 
then the gas is accelerated to the shock speed. 
The observed line profile is then a relatively narrow component centered at a velocity shifted from the
preshock gas by $v_s {\rm cos}(i)$. 
Figure~\ref{modfit} shows how such shocks could contribute to the H$_2$ line profiles.
It is particularly notable that clump B2 displays {\it only} what we interpret as 
J-type shock, while clump B1 is primarily CJ-type. We infer that the direction of the magnetic field relative to
blast wave changes significantly between B2 and B1.

\def\extra{
Not only is H$_2$ the most abundant species by far, but also 
emission from pure rotational H$_2$ lines is the dominant 
cooling mechanism of the gas. }

\def\extra{low velocity shocks; show H2 cooling is dominant compared to others, their Fig 4 REPEAT FOR OUR CONDITIONS;
 Also Fig5b showing shock density and temperature profile for J and C shocks; make new one for b=1, n=3000,  p=3e5 with boosts of 1(none), 10, 100 $\rightarrow$ vc=10, 32, 100
 \citep{lesaffre13}
}



\section{Discussion and Conclusions}

\subsection{Properties of shocks in IC~443}

There is a long history of spectroscopic observations of IC~443 and attempts to determine the nature of 
the shock fronts into them. We focused in this paper on addressing the question, initially posed in our
observing proposal: What type of shock produces the molecular emission from supernova remnants interacting with molecular clouds? 
We generated {\it a priori} predictions of the velocity profiles of the H$_2$ S(5) line, and compared them to new, SOFIA/EXES observations at high spatial ($2.4''$) and spectral ($R=7\times 10^4$) resolution.
We found that the CJ type shock into $10^3$ density and C-type into $10^4$ density are the best fit to most of the emission, with the high-velocity wings requiring J-type shocks into gas with
density $10^3$ cm$^{-3}$ and lower. 
In these shocks, the S(5) line of $H_2$ is 
predicted to be one of the brightest spectral lines and is therefore energetically significant as 
an important coolant in the shocked gas.

In comparison to previous work on the molecular shocks in IC~443:
\begin{itemize}
    \item \citet{richter95a} detected the H$_2$ S(2) line and compared with higher-excitation lines to infer a partially-dissociating
J-type shock into gas with density $4\times 10^3$ cm$^{-3}$; our results support the inference of a J-type shock based
on the detailed line profile and modern theoretical modeling.
\item \citet{cesarsky99} used {\it Infrared Space Observatory} to measure intensities of pure rotational lines
S(2) through S(8) for IC~443. Generally consistent with our results, they derived 
densities of order $10^4$ cm$^{-3}$ with shock speeds 30 km~s$^{-1}$ and ages 1000-2000 yr; indeed, their paper
contained one of the first applications of a non-steady CJ-shock.
\item \citet{snell05} observed H$_2$O at high velocity resolution using SWAS, showing emission, from a $4'$ region centered on clump B, with velocities up to 70 km~s$^{-1}$; 
they interpret the emission with a range of shock types and
infer properties of the oxygen state in the preshock gas. 
Our results can help explain the SWAS observations and the HCO$+$ and other broad molecular lines. The SWAS beam includes our position B1 and extends almost to B2, so the H$_2$ emission covers the entire
range of velocities where H$_2$O and HCO$^+$ were seen, and more. As for H$_2$, most of the broad-line
emission arises from CJ-shocks into dense gas, while the most extreme velocities are from J-shocks into
moderate-density gas.
\item \citet{rosado07} made moderate ($R=10^4$) resolution observations of the vibrationally excited H$_2$1--0 S(1) line toward IC~443C. 
The brightest emission in their image, was only marginally resolved but consistent with 
our much higher-resolution observation of the S(5) line. 
\end{itemize}

\subsection{Relationship between molecular cloud shocks and cosmic rays}

A remarkable achievement in high-energy astronomy was the first spatially-resolved image of a supernova
remnant, showing that TeV particles originate from the portion of IC~443 coinciding with shocked molecular gas
\citep{magic443,veritas443}. A joint analysis of the VERITAS and {\it Fermi} data documented the challenges
in combining the high energy data and yielded a map of the relativistic particle distribution \citep{humensky15}. 
In particular, the region around clump G stands out as a prominent TeV peak. 
The TeV photons could be produced by interaction of cosmic rays produced in the SN explosion with
a dense cloud, in which case the TeV photons are pion decay products \citep{torres08}.
By localizing the source of the high-energy $\gamma$ rays to the molecular interaction region 
(what we call the $\omega$ shaped ridge), as opposed to the northeastern rim where the radio emission 
is far brighter, an electron bremsstrahlung origin (in which the high-energy emission should arise
from the same place as the radio synchrotron emission) is ruled out, in favor of
a pion decay origin, meaning that localized high energy $gamma$ rays and TeV photons trace
the origin of (hadronic) cosmic rays \citep{tavani10,fermi443}. 
The prominence of clump G in TeV photons can be explained by that shock front both having dense gas
and being nearly edge-on, so that the path length is highest; it is also the brightest 
of the shocked clumps in 
the H$_2$ S(5) line, which is the brightest emission line from such shocks.

The general observational correspondence between the shocked molecular gas distribution and cosmic rays is only at the scale
of several arcminutes.
Our results on the H$_2$ emission from IC~443 show that the site of the shocks into dense gas, which may
be related to cosmic ray acceleration, are the portions of the molecular cloud with preshock densities
of order $10^3$ to $10^5$ cm$^{-3}$ with shock speeds of 20 to 60 \kms. The H$_2$ emitting region
tracing the shocks into dense gas is structured at the $2''$ scale (resolution limited) in regions with
size of order $30''$ (the size of the bright region in Clump C or G in the {\it Spitzer} image).
The angular resolution of the current state-of-the-art Cerenkov telescopes, including HESS, MAGIC, and VERITAS,
is approximately $5'$ for localizing 1 TeV particles entering the atmosphere.
The new state of the art will become the Cerenkov Telescope Array \citep{cta19}, which will have an angular
resolution approximately $3'$ at 1 TeV and $2'$ at 10 TeV. There is a fundamental limit to the Cerenkov
technique for localizing cosmic particles. Once all Cerenkov photons generated by particles
upon entering the atmosphere are detected, there is no further improvement of the centroids. 
This limit was estimated to be approximately $20''$ at 1 TeV and $10''$ at 10 TeV \citep{hofmann06}.
We expect that if the shocks into dense gas are intimately related with the comic ray origin, 
the structure in the TeV images will continue to be more detailed up until the ultimate limit of
the Cerenkov telescopes is reached.

It is likely that cosmic rays and dense molecular shocks are two integrally related phenomena, linked by 
the strength of the magnetic field. Cosmic ray acceleration requires a strong magnetic field; without a strong
field, the Lorentz force felt by high-energy particles will have a radius of curvature
(Larmor radius) so large that the particles will hardly be deflected before leaving the
region ($R>R_L$) for most astronomical objects. Magnetic field amplification is highest for fast shocks
into dense gas \citep{marcowith18}. As for the broad molecular line emission, we showed above that
strong (of order mG) magnetic fields are required, so that shocks driven by enough pressure to
accelerate the gas do not dissociate the molecules in the process. 
A relativistic particle of energy $10^{15}$ eV in a region with magnetic field strength 1 mG will
have a Larmor radius 0.001 pc, which is of the same order as the thickness predicted (in Fig.~\ref{fig:shthick})
for C-type shocks
into gas with the properties of the molecular cloud around IC~443.
Thus the common elements for cosmic ray
acceleration and broad molecular lines are a strong magnetic field and dense gas, which conditions
are both present when supernova-driven shock waves impinge upon molecular clouds.

\subsection{Future prospects}

The connection between cosmic ray origin and feedback of star formation into molecular clouds will make
understanding of supernova shocks into dense gas of continuing interest.
One curious observation that we can now answer is, why are broad molecular lines from shocks always
approximately 30 km~s$^{-1}$ wide. 
{\bfc Shocks faster than $\simgt 40$ \kms\ into gas with a low magnetic field will
be J-type and will destroy H$_2$ \citep{drd83}; 
given the emitted power increases steeply with shock velocity, we would expect molecular lines of $\simlt 40$ \kms width and
shifted relative to ambient gas.
}

Only with
very high velocity resolution are such effects discernible. The brightest, broad molecular lines are from
CJ-type shocks into dense gas, with magnetic field perpendicular to shock velocity. Unless the
magnetic field is always perpendicular to the shock velocity, there
should also be emission at more extreme velocities, due to gas that is shocked and not protected by the
magnetic field. In future theoretical work, it should be possible to unify J-shock models
\citep{hartigan87,hm89}, for which the immediate post-shock region is dominated by atomic physics
(ionization/recombination), with C-shock models \citep{godard19}, 
for which the post-shock gas is dominated by
MHD and chemistry.
Other theoretical work to better understand the shock physics involves simultaneous modeling of
the H$_2$ (and other molecule) excitation and line profile, as well as 2-D modeling of emission
from shocked clumps of shape more realistic than the 1-D slabs considered here \citep{tram18}.

\acknowledgements  
Based in part on observations made with the NASA/DLR Stratospheric Observatory for Infrared Astronomy (SOFIA). SOFIA is jointly operated by the Universities Space Research Association, Inc. (USRA), under NASA contract NNA17BF53C, and the Deutsches SOFIA Institut (DSI) under DLR contract 50 OK 0901 to the University of Stuttgart. 
EXES is supported by NASA agreement NNX13AI85A.
Financial support for the observational and theoretical work in this paper was provided by NASA through award \#06\_0001 issued by USRA.
\facility{SOFIA} 

\bibliographystyle{aasjournal}
\bibliography{wtrbib}

\end{document}